\documentclass[aps,prl]{revtex4}
\def\ee{\end{equation}}
\def\bea{\begin{eqnarray}}

\def\ul#1{ \underline{#1}}

\usepackage{graphicx}
\DeclareGraphicsExtensions{.png}
\begin{document}

\title{Semi-quantum Gravity and Testing Gravitational Bell Non-locality}
\author{ Adrian Kent}
\email{A.P.A.Kent@damtp.cam.ac.uk} 
\affiliation{Centre for Quantum Information and Foundations, DAMTP, Centre for
Mathematical Sciences,
University of Cambridge, Wilberforce Road, Cambridge, CB3 0WA, United Kingdom}
\affiliation{Perimeter Institute for Theoretical Physics, 31 Caroline Street
North, Waterloo, ON N2L 2Y5, Canada.}

\begin{abstract}
Semi-classical gravity attempts to define a 
hybrid theory in which a classical gravitational
field is coupled to a unitarily evolving quantum state.
Although semi-classical gravity is inconsistent with
observation, a viable theory of this type might be appealing,
since it potentially might preserve the basic features of our 
two most successful theories while unifying them.
It might also offer a natural solution to the quantum
measurement problem. 
I explore the scope for such ``semi-quantum'' hybrid theories, and note
some interesting, though daunting, constraints.  Consistency with observation
generally requires pyschophysical parallelism with the classical gravitational
field rather than the quantum matter.
Solvability suggests the gravitational
field at a point should be determined by physics in 
its past light cone, which requires local hidden
variables and predicts anomalously non-Newtonian
gravitational fields.   These predictions could
be tested by low energy, although technologically challenging,
experiments in which the Bell non-locality of the gravitational
field is verified by direct measurement.    
\end{abstract}
\maketitle
\section{Introduction}

Unifying general relativity and quantum theory remains one of 
the great unsolved problems of physics.   
Because the theories are based on radically different principles,
many approaches can be motivated, including the idea of 
a hybrid theory in which the gravitational field and
space-time are fundamentally classical, while matter
is quantum.   In principle there seems nothing inconsistent
in the idea of a hybrid theory.   Arguments for the necessity of quantizing
gravity have been made in the past \cite{eppley1977necessity},
but these have been refuted \cite{mattingly2006eppley,huggett2001quantize,albers2008measurement,
kent2018simple}.    

The path to a hybrid theory that has attracted most mainstream
attention to date is semi-classical gravity, in which a
classical gravitational field is coupled to the expectation
value of the quantum matter stress-energy tensor. 
Semi-classical gravity has not been shown 
to be consistently definable and is anyway empirically 
falsified  \cite{page1981indirect}.
Nonetheless, it has very interesting features.   
As generally understood, it is a proposal for
unifying unitary (Everettian) quantum theory with a classical
theory of gravity -- an extravagant hybrid more reminiscent 
of Greek mythology than zoology.   There seems to be 
general consensus that it has succeeded 
in being at least well enough defined to definitely fail: 
everybody accepts 
that semi-classical gravity predicts something \cite{page1981indirect} we do not see.   
I agree that it fails, in that there is no 
principled way of extracting predictions in agreement
with observation.   But I will argue that understanding
what it predicts involves a subtlety that seems to have
gone unnoticed and requires going beyond standard
discussions of Everettian quantum theory. 
This is because Everettian quantum theory coupled to classical gravity
raises a novel question about pyschophysical parallelism:
should or could it apply to the classical gravitational
degrees of freedom as well as, or even instead of, the quantum matter? 

Radical though it this to think of the structure of space-time 
as a (or even the) substrate for a lawlike description of consciousness, it
seems a plausibly consistent possibility that deserves some attention.
A hybrid theory could, in principle, be consistent with observation 
on this hypothesis, while failing on the more standard hypothesis
that psychophysical parallelism applies (only) to quantum matter.  
In principle, a hybrid theory that is empirically successful in this sense
might give an elegant solution
to the quantum measurement problem without requiring 
either collapses or many worlds containing effectively
independent observers: in such a theory, space-time could play the
dual role of being the arena for and the subject of 
quantum theory.  
Even if there turns out to be no promising way of developing such
theories, it would be good to establish this. 

Semi-classical gravity's prediction that even in an otherwise Newtonian regime
the gravitational fields need not be sourced by 
the observable matter is also intriguing.
The specific anomalies predicted by semi-classical gravity
are empirically refuted, of course.
However, the predictions themselves are not
{\it logically} inconsistent, and indeed it may yet turn out that
semi-classical gravity itself is a fully consistent theory.   
It seems natural to ask whether 
there might be any other hybrid theories predicting  
subtler gravitational anomalies that might
not yet have been tested. 

These seem good reasons to explore the possibility
of defining other hybrid theories that respect quantum
unitarity.   Unsurprisingly, as with semi-classical gravity,
there are conceptual difficulties. 
The easiest class of theories
to define seem to be those in which the gravitational field
at a point $P$ depends only on physics in its past light
cone $\Lambda(P)$.   These need a local hidden
variable to define the gravitational field, which reduces their conceptual elegance.   
They are, however, testable, and
motivate an interesting technological challenge for
experimenters. 

\section{Semi-classical gravity}

Semi-classical gravity is an interesting  
theoretical proposal that illustrates some of the difficulties
in defining hybrid theories of classical gravity and quantum
matter.
Setting
\begin{equation}
\label{scg}
G_{\mu \nu} = \langle T_{\mu \nu} \rangle  \, , 
\end{equation}
for some appropriately defined expectation value
of the quantum matter stress-energy tensor,
gives, formally, an equation that links a classical tensor
defined by the gravitational field to a classical tensor 
derived from the quantum matter state.   

In principle, semi-classical gravity may be defined
for any version of quantum theory for which the
local stress-energy tensor is well-defined. 
For example, it could be defined for theories
with localized collapses that only affect the
quantum matter state, and hence $T_{\mu \nu}$,
in their future light cones \cite{kent2005nonlinearity}.   

In practice, it has most often been considered in
an Everettian model in which unitarily evolving quantum matter
is coupled to gravity via Eqn. (\ref{scg}). 
This raises its own issues, since there is no 
consensus on how to understand Everettian
quantum theory \cite{saunders2010many}.  The problems of understanding
probability and branch structure in Everettian
quantum theory are problematically intertwined, as is the 
question of whether any version of the theory
is testable \cite{saunders2010many}.      

That said, the idea of coupling 
a unitarily evolving quantum state to a classical
model of space-time also raises
intriguing new possibilities.
One of these is that adding the classical space-time 
could solve the problems of Everettian quantum
theory -- or, to put it more neutrally, could 
offer a radical alternative perspective on that theory.
One of the aims of this paper is to explore the
implications of this idea.    Another is to 
consider the Page-Geilker experiment, which was
explicitly motivated by an Everettian version
of semi-classical gravity.   This also motivates
further extensions of that experiment in order to test the Bell
non-locality of the Newtonian gravitational field.   So I will follow the mainstream tradition 
here, and assume that quantum matter follows a unitary
evolution law, except where otherwise stated.

The natural interpretation of the right hand side of (\ref{scg})
is as an expectation value calculated in a given background
space-time.
However, this requires regularization, and 
also affects the dynamics of space-time through
the left hand side. 
This makes the back-reaction problem both
technically and conceptually challenging (see
e.g. \cite{hu1995back,flanagan1996does,martin1999stochastic,giulini1995consistency,hu2000fluctuations} 
and references therein).     
Whether stable weak field solutions of semi-classical
gravity even exist has been questioned
\cite{horowitz1980semiclassical}. 
Its status as a complete theory is thus at best conceptually
and technically unclear.   

Semi-classical gravity's technical and 
conceptual problems may yet prove to be resolvable.
However, it fails empirically.  
In the familiar regime of non-relativistic quantum mechanics
in weak gravitational fields, semi-classical gravity makes
the counter-intuitive prediction that a macroscopic mass
put into a superposition of macroscopically distinct position
states creates the same gravitational field -- an average of
the fields associated with the two mass states -- in both
branches.   
In an Everettian framework, this can be tested by 
using a quantum experiment to create two Everett branches
and arranging different macroscopic mass distributions
in the branches by moving masses in a way that depends
on the quantum measurement outcome. 
All Everettians agree, at least, that a human
observer after the experiment should see a world
characterized by one of the two branches, while
the full quantum state  includes a superposition of both. 

The result was 
thought uncertain enough by Page and Geilker \cite{page1981indirect}
to motivate their  
well-known experiment, which confirmed the general expectation
that the field is sourced by the observed mass state rather than
by an average of the observed and unobserved states.\footnote{
The Page-Geilker experiment is sometimes discussed as though
it were an elaborate joke, but it was genuinely intended as
an empirical test of an uncertain proposition.\cite{pagepriv}}

As far as I understand it, almost everybody was (or should have been)
almost persuaded that semi-classical gravity was incorrect before
the Page-Geilker experiment, 
because it is extremely hard to understand how semi-classical
gravity could be true locally without implying that we should also 
see gravitational fields averaged over many different
cosmological outcomes, and so not at all closely 
correlated with the matter distribution in our own
branch.   
Page and Geilker nonetheless regarded this as
plausible, although unlikely \cite{page1981indirect}. 
One might, perhaps, think it plausible 
if one takes seriously the possibility that the
cosmological initial state could have been fine-tuned to 
produce the large-scale distribution of matter
that we see, in almost all Everettian branches.   
Alternatively, one might appeal to the possibility that
the nonlinearities and other complexities of semi-classical gravity
somehow lead to a definition of consistent solution
that tends to imply alignment of the quantum matter with the gravitational
field on large scales. 
Neither line of thought seems easy to defend, though.
In any case, their experiment \cite{page1981indirect} removed
whatever slivers of doubt remained, and persuaded everybody.  

Clearly, one can argue about 
reasonable Bayesian priors for the Page-Geilker experiment. 
Still, the fact that it is still discussed at quantum gravity
meetings decades later suggests it was justified. 
For me, the broader scientific moral is that we should not reject a
classical gravity quantum matter hybrid model simply because it predicts
non-Newtonian correlations between the matter and the gravitational
field in (otherwise) Newtonian regimes, or more generally a 
failure of Einstein's equations.
\footnote{Although there is no evident connection, the dark energy problem
probably reinforces this moral a little.  Some might also
include the dark matter problem.}
This would be counter-intuitive, of course, but
there are surprises in physics, and the boundaries between
quantum theory and gravity seem a likelier regime than most
in which to find them.  

\section{Semi-quantum gravity theories}

Given a superposition of macroscopically
distinct states, semi-classical gravity takes the average mass
distribution.  
Although this fails empirically, semi-classical gravity has 
some interesting features.   This motivates exploring other possible
hybrid theories that combine unitarily evolving quantum matter
with classical gravity: let us call them {\it semi-quantum gravity}
theories. 

Now, a full Everettian quantum theory of gravity,
given a superposition of macroscopically distinct states, predicts it will extend to
a superposition of spacetimes, entangling gravity and matter.
This may be correct, but does not give what we want here -- a model of classical gravity 
interacting with quantum matter.  

What are the other options?   Cosmological observation and 
the Page-Geilker experiment suggest that a theory with
a classical space-time should -- at least in  
contexts tested to date -- predict the gravitational
field to behave as though sourced by one random element from the
superposition.    If this is supposed to collapse the superposition, then
we no longer have unitary quantum evolution.   This line
of thought leads in the direction of gravitational 
collapse models, for example of the types originally 
proposed by Diosi \cite{diosi1987universal} and 
Penrose \cite{penrose1996gravity}.  These are 
very well motivated and interesting proposals, which
are actively being developed (see e.g. \cite{tilloy2016sourcing}), 
but, again, not the type of hybrid theory we want to explore here.  

So, we would need to assume that the gravitational field is effectively
sourced by {\it one Everettian branch}, while the quantum matter 
{\it continues to evolve unitarily}.   We next review the possibilities
and problems this raises. 

\section{Semi-quantum gravity theories and conscious perception}

If the gravitational field is sourced by one Everettian branch, while
the quantum matter continues to evolve unitarily, then 
every Everettian branch in the superposition
is associated with the {\it same}
classical gravitational field, since by hypothesis 
that field defines a single definite space-time.    
This seems to imply that after any given run of the Page-Geilker
experiment, observers will see a definite gravitational field,
corresponding to one of the two possible configurations, 
{\it but not necessarily the configuration realised in their
branch}.   Similarly, it seems to imply that astronomers will see definite gravitational
fields corresponding to a definite configuration of celestial
bodies, but typically it will not resemble the configuration
realised in their branch.  Obviously, this is not our experience. 

This leaves two logical possibilities to consider.

First, suppose that we want to maintain the standard Everettian
view that all branches of the wave function potentially contain
conscious observers, and that many quasiclassical branches similar but
not identical to ours actually contain conscious observers similar
to ourselves.   Suppose also, as most Everettians do,
that some sort of probabilistic reasoning about our 
expected observations
can somehow be justified in this picture.    Everettians disagree on how
probabilities are supposed to emerge, and it is hard to 
find language consistent with every Everettian proposal here.  
But, very roughly speaking, we are meant to conclude that if ``we'' are 
represented in two different branches,  the
probability ratio of ``finding ourselves'' in the respective branches
is the Born weight ratio.\footnote{Everettian readers are
encouraged to reword this according to their preferred account
of Everettian probability.}    Suppose also that the gravitational
field -- acting on all the branches -- behaves as though sourced
by the mass distribution in one of them.    
Then, for consistency with observation, we would have to be 
able to argue that only the ``sourcing'' branch is able to 
support conscious observers, or at least that typical conscious 
observers belong to a branch that is hard to distinguish from
the sourcing branch.   This is only possible if something stops
observers from existing or consciously functioning in the remaining branches.    

Now, from a cosmological perspective, some statements
in this direction seem justifiable.   For example, consider a branch 
in which  the matter from which the
solar system formed was in one region of space-time, but 
acts under the influence of gravitational fields derived from a branch
in which the matter was distributed very differently. 
Our sun and planets almost certainly would not have formed in this branch,
and so we would not have evolved.   More generally, wild large scale discrepancies 
between matter configurations and gravitational fields are 
likely not conducive to life, still less to conscious
and scientifically inclined observers.   

For terrestrial experiments and other smaller scale 
interactions between quantum matter and the gravitational
field, though, this line of argument seems doomed.
In a Page-Geilker experiment, while it would certainly be disconcerting
to put a metal ball in one place and see the gravitational field
behave as though it were in another, it 
is hard to see any reason why it should stop us consciously functioning.
It is true that, in that branch, our brain would also be effectively in a state in
which its matter distribution is in one brain state (that of having
seen the experiment with outcome $\alpha$) and its gravitational
field would correspond to another brain state (that of having
seen the experiment with outcome $\beta$).   The difference in
the corresponding mass distributions is tiny, but perhaps
large enough that the gravitational fields are in principle
distinguishable.\footnote{For some relevant discussions, see
e.g. Refs. \cite{bassi2010breaking,kent2018perception}.}
However, the self-interaction between the brain's matter
state and its gravitational field is small, and 
its effects on the brain seem to be utterly negligible in any 
plausible model of brain function.  

This leaves the second option, which highlights a
seemingly strange but intriguing and 
fundamental question.   This is whether consciousness
might be associated with the classical gravitational degrees 
of freedom rather than the quantum matter degrees of freedom.
This question perhaps makes most sense to those willing to entertain the hypothesis that 
there plausibly could be a 
physical law-like connection between conscious perceptions 
and material physics, and specifically that it is logically
conceivable \cite{sep-zombies} that there could be a law that 
implies that one physical system is conscious and 
another is not, even if they function similarly in terms of
information processing.    In particular, it is unlikely 
to seem sensible to Everettian functionalists \cite{sep-functionalism}, who 
believe that it essentially follows from the 
definition of consciousness that it must be 
associated with any set
of variables approximately describable within a 
branch as behaving like an observer interacting
with an environment, and so in particular that
there must be conscious observers in every
quasiclassical branch arising from a quantum
experiment.\footnote{I don't have any new
argument against functionalism here,
so any dogmatically functionalist readers may
want to stop at this point.}

Even if one accepts there could be a law-like theory
of consciousness, it may still initially seem absurd to suggest that
such a law could associate consciousness
with gravity rather than matter. 
After all, we understand our evolution in terms of biochemistry 
and quasiclassical equations of motion that are 
underpinned by quantum theory, and we understand the
brain as an information processing system that 
appears to be classically modellable and again 
is underpinned by quantum theory. 

So we do, but it does not follow that any law-like theory
of consciousness {\it has} to associate consciousness
with quantum matter. 
For example, in domains where classical general relativity
(with classical matter sources) is a good effective theory,
it makes no sense to say that the right hand side of
$G_{\mu \nu} = T_{\mu \nu}$ can explain or characterize emergent 
phenomena and the left hand side can not.  
In principle, all the relevant parts of evolutionary theory
and of neuroscience have parallel descriptions in 
the gravitational degrees of freedom.   
The gravitational fields associated with neural
signals are admittedly very weak indeed.
It is true that in some models they may not be detectable even in
principle, and in any such models, consciousness indeed cannot sensibly
be associated with gravity in a lawlike way.\footnote{
Again see e.g. Refs \cite{bassi2010breaking,kent2018perception} for
some related discussions.}
Still, we can not assume this a priori: in the
present state of our understanding it seems quite possible
that different conscious mind states are normally associated
with brain states whose gravitational fields are detectably different
in principle.  

Given that, and with the caveats noted, 
we have a possible first line of defence of the type of 
hybrid theory under discussion.   Many branches are 
represented in the quantum matter superposition, but
only one of them is effectively a source for 
the classical gravitational field.   Conscious observers 
are conscious by virtue of the properties of the classical
gravitational field, and in particular the gravitational
fields associated with their brain states. 
Their observations of the matter and gravitational 
degrees of freedom therefore both correspond to the
quasiclassical physics described by the same selected branch.

Even if this defence is accepted, there are more
problems to address.  
We have been vague so far about whether the branch selection
should be global (one branch is selected from the universal 
wave function) or local (the gravitational field
at different points in space time may be sourced by matter distributions belonging to
different branches) and about the form of any branch selection rules.
We will turn to these questions below. 
We first reexamine semi-classical gravity in the light of this
discussion.

\section{Conscious perception and semi-classical gravity}

First consider a standard Everettian account of a simplified Page-Geilker
experiment, based on quantum gravity intuitions.
We suppose for the moment that the experiment involves manual
interventions by the experimenter, who moves masses in response to
Geiger counter detector readings and then measures the associated
gravitational field in a Cavendish experiment.  
There are effectively two branches, which we take to be equiprobable.
These correspond to two possible outcomes
of a quantum experiment with a radioactive source and a Geiger counter
detector.   In one branch, an experimenter observes 
outcome $\alpha$, moves a mass to position A, measure the gravitational field,
and observes results $A_g$.   In the other, he observes $\beta$, moves
the mass to $B$, and observes $B_g$.   

At the quantum level, both the relevant observations involve 
light scattering from dial readings, reaching the experimenter's  
eye, causing biochemical processes in rod cells, and 
causing neurons to fire in his brain.  On an Everettian
view, all of these are 
considered as quantum systems.    

In particular, it is
ultimately the influence of the classical gravitational
field on quantum matter that allows the experimenters
in each branch to become aware of the gravitational
field in their branch, and to verify that it behaves as
though sourced by matter in their branch.

A standard Everettian account of the same experiment
based on semi-classical gravity runs slightly differently.
In both branches the experimenter observes the same gravitational
field, an average $M_g = \frac{1}{2} (A_g + B_g )$.   In principle, the relevant parts of his
brain processes could have the same quantum description.
In practice, this seems unlikely, since the brain is 
not a simple deterministic machine, and his brain
is preconditioned differently by the two different
observations of the quantum experiment and by moving
the masses.   

We could make it somewhat more plausible in an
automated version of the experiment, in which
he never observes the quantum outcome ($\alpha$
or $\beta$) or the moving masses, and in which
the gravitational field measurements 
is announced at a
precise predetermined time, independent
of the rest of the experiment.   
The experiment would need to be very well screened
from the experimenter, so that for example the 
relevant parts of his brain are not affected differently
by electrostatic charges in the two branches.
Let us suppose this is possible. 

In either the manual or the automated experiment, we have
potentially at least three generalized ``branches'' to which
psychophysical parallelism could be applied. 
Two of these are the standard Everettian  branches
defined by the quantum matter.   In these the observer
has well-defined brain states, corresponding to 
having observed the relevant outcome of the 
Page-Geilker experiment (in the manual version of the experiment) and to 
having observed the gravitational field corresponding
to the average of these two branches.   
The third is defined by the gravitational field.
In this the observer's brain has a gravitational
field defined by the average of the gravitational fields corresponding
to the matter distributions in the two Everettian branches.   

In the manual version of the experiment, this is not 
a direct encoding of either outcome.   If one applies
psychophysical parallelism to it at all, one might initially
be tempted to interpret it as ``confusion'' or ``no 
conscious perception of the Geiger counter reading''.   
Thinking further, one might decide that a law of 
psychophysical parallelism {\it could} interpret
a gravitational field of the form
$$
\sum_i p_i \Phi_i (t) \, , 
$$
where the $\Phi_i (t)$ are evolving gravitational 
fields that represent coherent pictures
of the gravitational field associated with a functioning conscious brain, 
as a set of probabilistically weighted branches,
with an observer having probability $p_i$
of ``finding himself'' in the branch described
by field $\Phi_i$.    

This may seem
strange, even desperate.   But our understanding of psychophysical
parallelism is murky.   It is not so clear that
it is stranger than the Everettian view that 
a quantum state of the form
$$ 
\sum a_i \Psi_i (t) \, ,      
$$
where the $\Psi_i (t)$ are evolving quantum  
states that represent coherent pictures
of the matter of a functioning conscious brain, 
as a set of probabilistically weighted branches,
with an observer having probability $| a_i |^2$
of ``finding himself'' in the branch described
by state $\Psi_i$.   Both ultimately rely on the 
fact that the full state can be represented 
mathematically as the sum of states that are
quasiclassical and so (in some approximation) 
have a compressible description.  
Of course, the sums are in different spaces, and 
different algorithms are used to extract the relevant components.
These differences {\it could} be significant, on some possible views of 
psychophysical parallelism, but my sense is that neither 
Everettians nor others have elaborated a coherent enough
account of pyschophysical parallelism to make a compelling
argument that they are {\it necessarily} significant. 

In the automated version of the experiment, one might
initially interpret the gravitational field
as defining a conscious state of ``no consciousness of (or confusion about)
the Geiger counter reading'' combined with ``certainty
about the Cavendish experiment measurement of the
gravitational field''.   Thinking further, as above,
one might decide to reinterpret this as two
branches that give different but definite perceptions
of the Geiger counter reading and that merge to 
give the same perception of the Cavendish experiment
measurement.  

Whichever version of psychophysical parallelism one 
applies to the gravitational
field, it does not rescue semi-classical gravity:
none of the mathematical structures that could 
plausibly be considered ``branches'' describe
an experience in agreement with observation.
We either replicate the Everettian account,
or we add an account even more at variance
with observation.    In this sense, there is no positive gain
from applying psychophysical parallelism to 
both gravity and matter in semi-classical gravity:
it does not make the theory better and it may make
it worse.    But this does not mean it is the
wrong way of interpreting the theory; it simply
means the theory is wrong.   
To decide what the ``right'' form of psychophysical
parallelism is in this or any other theory, we
would need some general meta-principle of pyscho-physical
parallelism applicable to very general physical theories 
(perhaps including theories that contain neither matter
nor space-time in any standard sense), and then
apply this meta-principle to the relevant 
theory.
Semi-classical gravity teaches us that when a theory 
includes gravitational and matter fields that follow
different laws, there is more than one plausible
option for psychophysical parallelism.\footnote{Some physicists appear to believe
there is no theoretical mystery about psychophysical parallelism, 
that it is evidently and necessarily applied to matter only.
Some also believe that it is evidently and necessarily Everettian.   
It {\it could} be a fact about the world that psychophysical
parallelism applies to matter only, and it could even be a 
fact about the world that it is essentially Everettian.
But it seems to me \cite{kent2010one,kent2016quanta} these could only be contingent,
rather than necessary, facts about our universe.}

\section{Can hybrid models reproduce quantum gravity intuitions?}   

\subsection{ Quantum gravity intuitions}

One way of capturing Everettian intuitions about quantum gravity
is that some sort of effective branching emerges from a
generalized path integral.   Many of the branches
effectively characterise structure forming in
the universe, with the matter and spacetime then evolving quasiclassically.
There are a very large number of such quasiclassical
regimes emerging from different branches.
These may differ in the details of the structure on all scales -- 
from the distribution of galaxies to the outcomes of 
small-scale terrestrial experiments.
It is an open question whether they typically
follow essentially the same quasiclassical laws,
or whether these may also vary widely.
A good theory ought ideally to predict that
the quasiclassical world we observe is in
some suitable sense typical, or at least typical
conditioned on the existence of observers.

Quasiclassical regimes may contain
regions in which space-time is approximately
flat, with an inertial
frame in which macroscopic matter generally
moves at non-relativistic speeds,   
and where Newtonian gravity is a good 
approximation.  Our neighbourhood is an example.
Everettian quantum gravity intuitions 
imply that, in such regions, the Newtonian gravitational field 
should appear be sourced by the observed mass distribution.   

\subsection{An obstacle to defining hybrid models via global branch selection}

\begin{figure}[h]
\centering
\includegraphics[width=\linewidth, height=10cm]{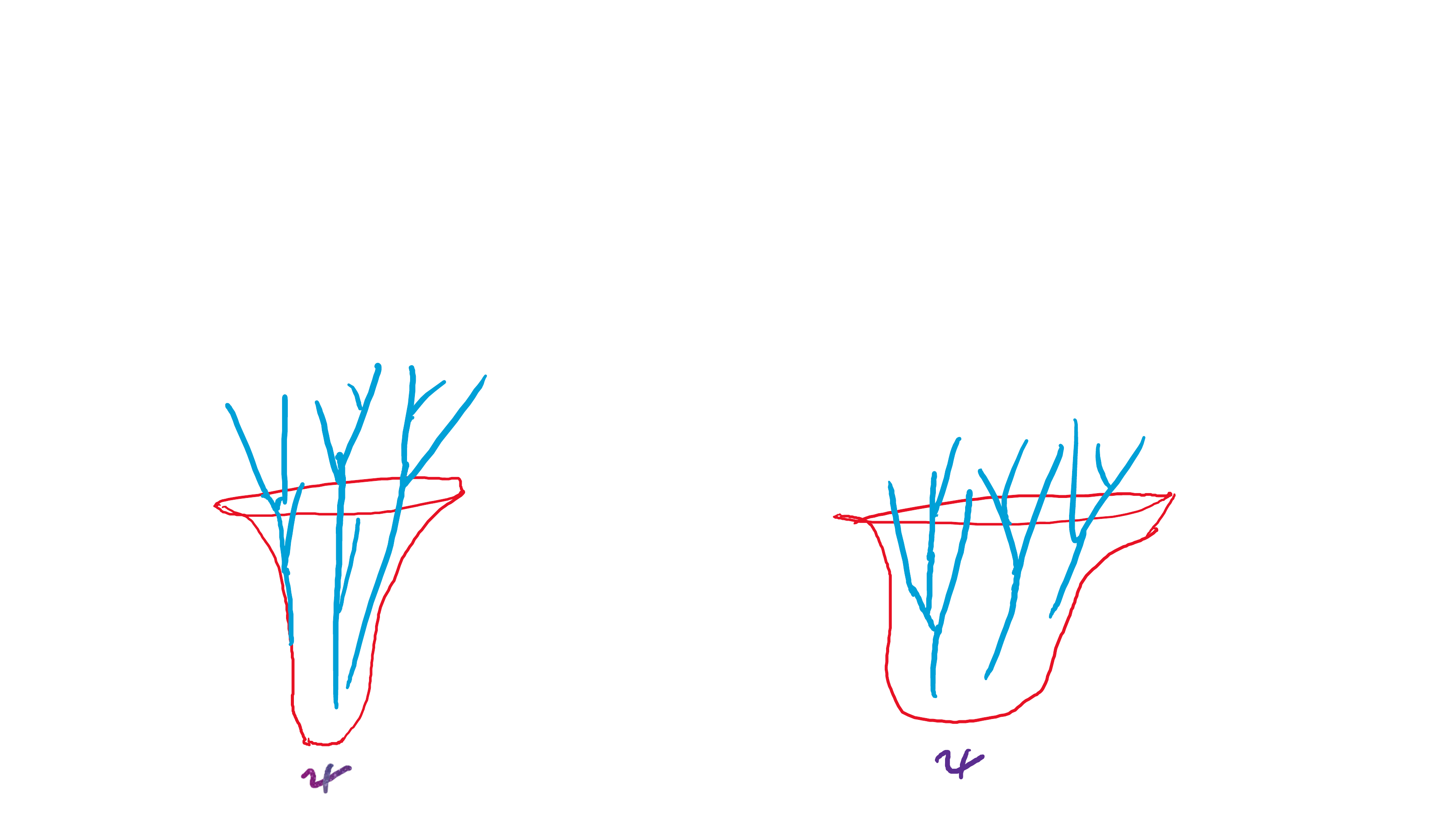} 
\caption{For a given initial state, the Everettian branching
structure depends on the classical space-time.   Postulating 
laws in which the classical space-time depends on the Everettian
branching structure creates a global consistency problem.}
\label{evgravbranch}
\end{figure}

In a hybrid model, we can only apply Everettian intuitions to 
the quantum matter.   Whatever rule defines the classical spacetime,
it gives a fixed arena in which the quantum matter evolves.
The Everettian branching structure will thus be radically different
from one that would emerge from a full quantum gravity theory:
we can {\it not} approximate it simply by ignoring the gravitational
degrees of freedom.   (See Fig. \ref{evgravbranch}.)

This highlights a difficulty in defining
hybrid models by a global rule based on Everettian branches.
We cannot naively determine a classical space-time by selecting
a global Everettian branch, because the Everettian branches
depend on the classical space-time. 

As in the case of semi-classical gravity, we have a consistency
problem.  There is a significant difference, though. 
In semi-classical gravity, we can (at least formally) 
clearly understand what constitutes a consistent solution,
namely a background space-time and initial conditions
on and from which the unitary quantum
matter field evolves so that its expectation value satisfies
(\ref{scg}).   However, we do not expect to find any
experimentally or cosmologically relevant background space-times $S$  
and initial states $\psi_0$ with the property that every Everettian branch 
arising from $\psi_0$ in $S$ generates $S$ when considered as the
matter source: the Page-Geilker experiment precludes this. 
A ``consistent implementation'' of our global branch selection
postulate seems to require something like a sample space
$\{ B_{\lambda} , S_{\lambda} \}_{\lambda \in \Lambda}$ with
probability measure $\mu$,
where each $B_{\lambda}$ is an Everettian branch from the
branching structure $E_{\lambda}$ defined by the background space-time
$S_{\lambda}$, and the $E_{\lambda}$ are generally all different.  
It is not at all clear either how to arrive at such a sample
space and measure, or how to define what we mean by ``consistent''
here.   Unless and until there is a good proposal for resolving these conceptual
issues, the idea of defining hybrid models by a global branch
selection rule seems a non-starter.  

This leaves the possibility of using
local branch selection rules to define
hybrid models, and accepting that any
such models will disagree with generic predictions
based on Everettian quantum gravity intuitions. 

\section{Local branch rules}

A local branch rule implies that the gravitational field
at a point $P$ is sourced by the mass distribution obtained 
in the neighbourhood of $P$ from some Everettian branch
$B(P)$.   As the notation implies, this branch choice in general
depends on $P$, as may the Everettian branching structure $E(P) \ni
B(P)$. 
The rule needs to give an algorithm for obtaining $B(P)$. 
Since there is no obvious
candidate for a deterministic local rule, we suppose $B(P)$ is
chosen probabilistically.    
To explain why the gravitational field generally appears to 
be sourced by matter in the standard way, we assume, as above, that
psychophysical parallelism is with the gravitational field. 
To explain the appearance of
the Born rule in quantum experiments, we assume that a branch
$B(P)$ is chosen with probability given by its Born weight.

\begin{figure}[h]
\centering
\includegraphics[width=\linewidth, height=10cm]{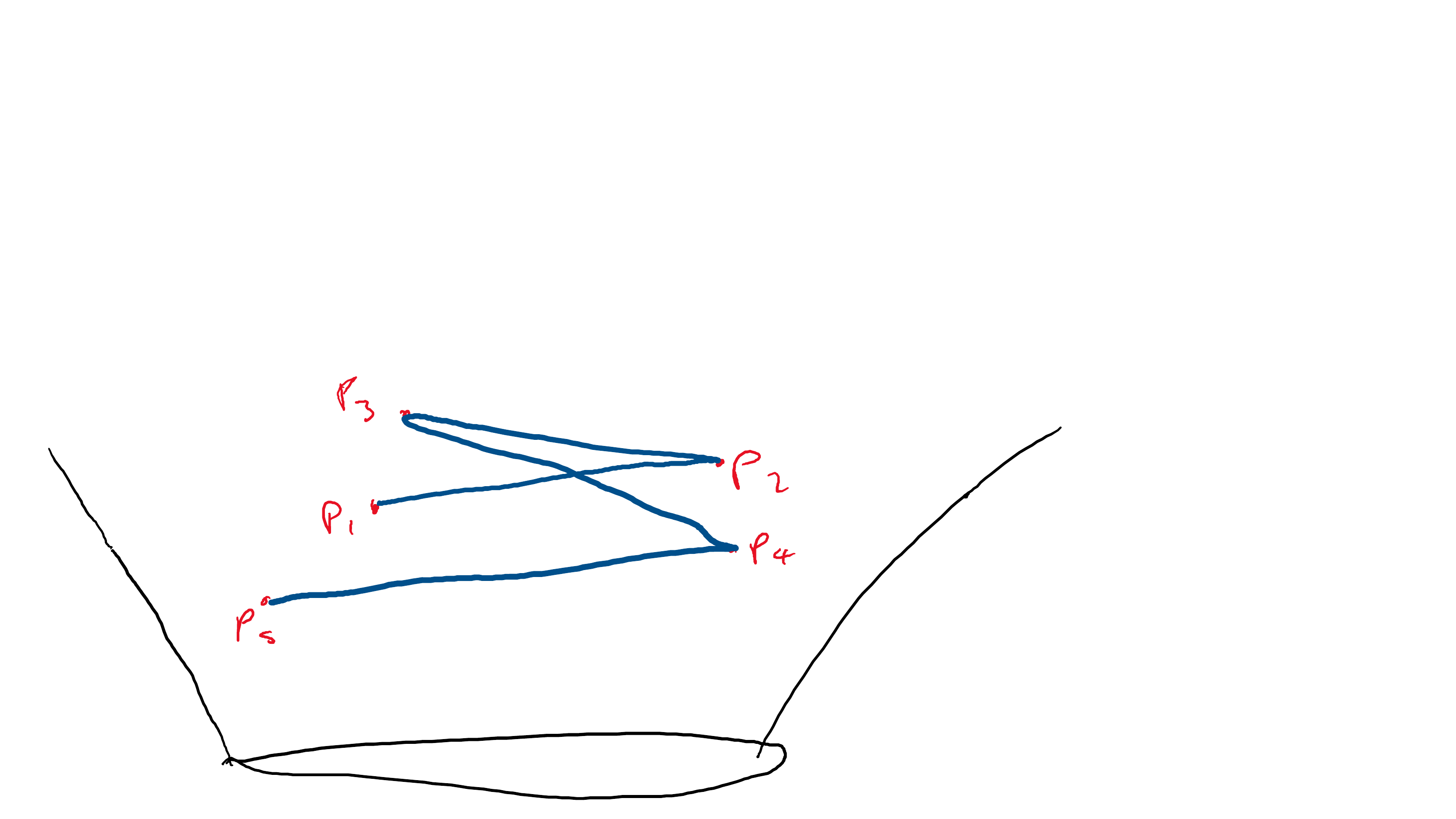} 
\caption{If the branch choices and hence gravitational fields at $P_i$
are non-locally correlated with those at the spacelike separated point
$P_{i+1}$, for each $i$, the global structure of space-time may depend
on branch choices throughout space-time, while the 
Everettian branching structure depends on the global structure of
space-time.  We thus again have a global consistency problem.}
\label{spacelikeconsistency}
\end{figure}

Now, if the branch choices at space-like separated points $P$ and $Q$ 
may in general be non-locally correlated, a consistent solution
may have to respect consistency conditions that apply throughout
space-time.   (See Fig. \ref{spacelikeconsistency}.) 
We again reach the conceptual obstacle that 
the branching structure depends on the space-time, 
the space-time depends globally on the branching structure,
and it is not immediately clear how to define, let alone find,
consistent solutions. 

We can avoid this particular difficulty by assuming that the branch choice, 
and hence gravitational field,  
at $P$ depends only on physics in its past light cone $\Lambda (P)$.
The gravitational field, and hence structure of space-time in $\Lambda
(P)$ is already defined by earlier branch choices.
A hybrid theory of this type may and presumably should
have its own gravitational dynamics, which imply that the
gravitational field (or more generally metric) at $P$ also depends on the 
structure of space-time within $\Lambda(P)$.  
Unless we assume the dynamics obey a suitable
differential equation or Markovian stochastic equation,
the dependence could in principle be on the field throughout $\Lambda
(P)$.   However, for the theory to connect matter, gravity and
observation sensibly we need to assume that, whatever the 
nature of this dependence, the observed gravitational field in regions
around space-time points $P$ does indeed normally behave, to very good
approximation, as though sourced from some Everettian branch $B(P)$,
chosen randomly according to Born weight from an Everettian branch
structure $E(P)$ that is defined by the matter states and space-time
in $\Lambda(P)$. 

We also need that the branch choice rule usually ensures
some sort of continuity in the gravitational field,
for two reasons.  First, we would like the gravitational
field's evolution to define a generally consistent narrative
of quasiclassical physics.  
There is a logically consistent alternative,
in which $B(P)$ is independently randomly chosen. 
But, as Bell persuasively argued \cite{bell1995quantum}, it is hard to take seriously
a world comprising uncorrelated momentary events, in which
records and memories appear to give evidence of histories that in
fact never took place.    
Second, we would like a generally continuous space-time to
emerge, to allow the possibility
that quantum evolution
and a branching structure might 
be well defined.\footnote{We should add the caveat that without a precise
theory, which we don't currently have, these cannot be guaranteed.}

Effectively, then, we are assuming that $B(P)$ is determined
by a {\it locally causal} hidden variable that is
determined by physics within $\Lambda(P)$.  
We are also assuming that there are rules determining the gravitational
field around $P$ from physics within $\Lambda(P)$, and that these
allow a well-defined unitary evolution of the quantum state within
$\Lambda (P)$ and a well-defined Everettian branching
structure $E(P)$, and so a well-defined
sample space and probability distribution for $B(P)$. 
To be clear: neither Everettian quantum theory nor classical general
relativity imply the existence of rules that have
all the desired properties.
To continue the discussion at present we just have to assume
some such rules exist.  

\subsection{Alignment with intuitions from general relativity}

Another (related) way of motivating probabilistic branch
dependence and the introduction of
local hidden variables is to start from 
the principles of general relativity.
From this perspective, it seems natural 
to consider local hybrid models in which
the gravitational field respects 
the strict form of Einstein causality 
embodied in general relativity, in 
which the metric (or local gravitational
field) at any point $P$ depends only on the
metric (or field) and matter states in its 
past light cone $\Lambda(P)$. 

Now, in Everettian quantum theory in a classical
space-time, matter states are defined on spacelike 
hypersurfaces.   On any hypersurface $S$ through
a point $P$ in our region of space-time, the 
matter state will be a massively entangled
superposition, and the reduced density matrix
at $P$ a high entropy mixture.   Neither
description gives a good candidate source term
for the gravitational field: both imply the
same incorrect averaging prescription given by
semi-classical gravity for the Page-Geilker
experiment.   

We thus have to 
suppose that the hybrid model laws 
are probabilistic, so that some
hidden variable determines $B(P)$ 
as above.   If the gravitational field at $P$
depends only on physics in $\Lambda (P)$,
then this has to be a locally causal hidden variable. 
This constrains the possible predictions of local hybrid models,
as we now discuss.    

\subsection{Local hybrid models and the Page-Geilker experiment}

In Page and Geilker's experiment \cite{page1981indirect}, $\gamma$ rays from a cobalt-$60$
source were detected over $30$ sec intervals by two nearby
Geiger counters, which detected on average $1509$ and $888$
counts respectively.   Runs in which the ratio 
of detections was higher or lower than a given value, chosen
so that the outcomes were roughly equiprobable, were 
assigned overall outcome $\alpha$ or $\beta$ respectively, 
and masses in a Cavendish experiment were placed in
the corresponding one of two possible sequences of configurations.
The gravitational field was then measured over $30$ min, and
found to agree with Newtonian predictions and disagree with
those of semi-classical gravity.  

\begin{figure}[h]
\centering
\includegraphics[width=\linewidth, height=10cm]{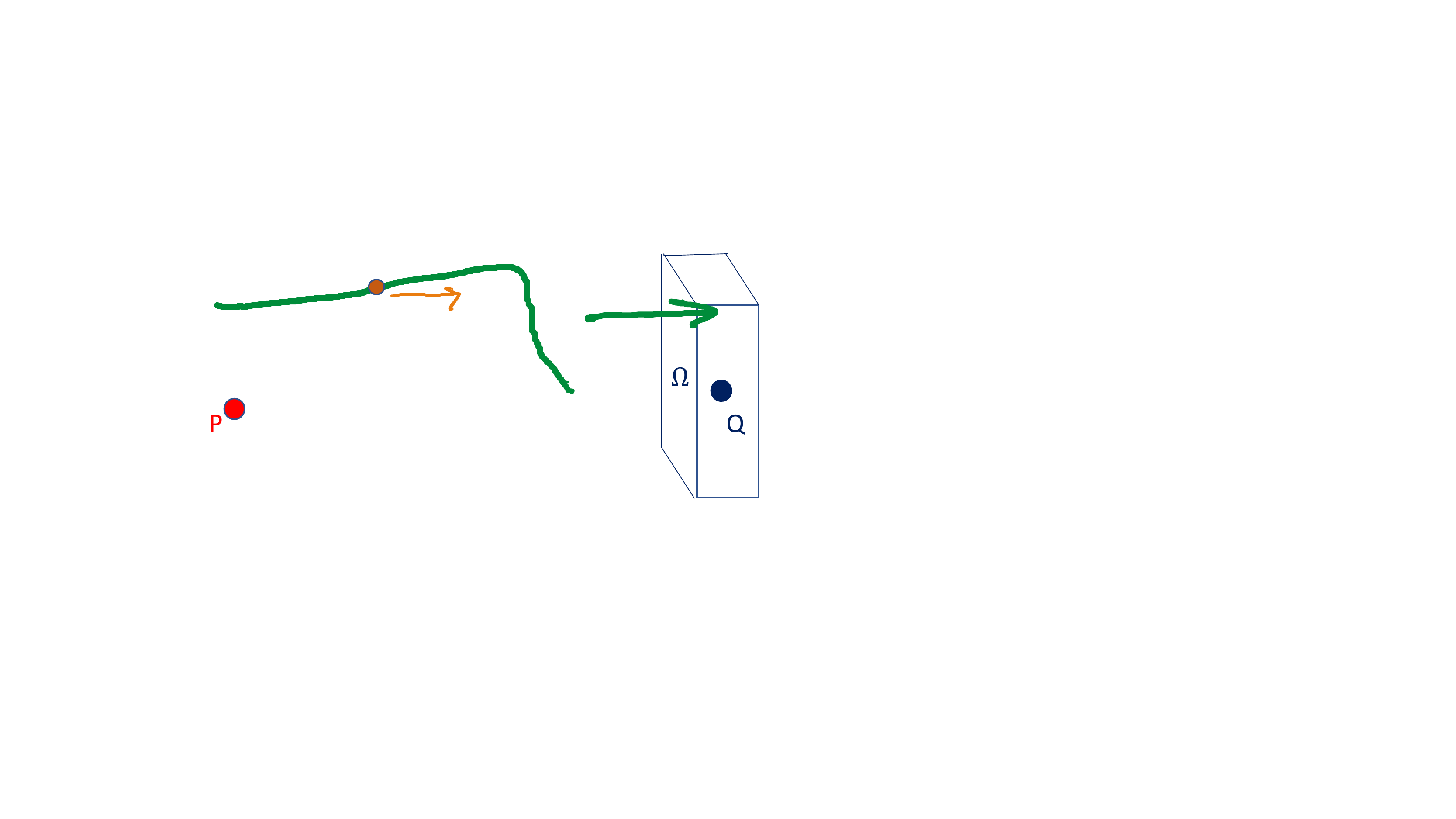} 
\caption{Graphic illustration of a simple local hidden variable model for nuclear
decay.  A ``hidden particle'' associated with the emitted photon
wave function is produced by the nucleus at a specified emission time
and propagates with a specified velocity from the source at $P$.
The time and velocity are 
local hidden variables in this model.  
A detector, centred at $Q$, subtending solid angle $\Omega$, clicks if and when the ``hidden
particle'' reaches it.}
\label{decay2}
\end{figure}

\begin{figure}[h]
\centering
\includegraphics[width=\linewidth, height=10cm]{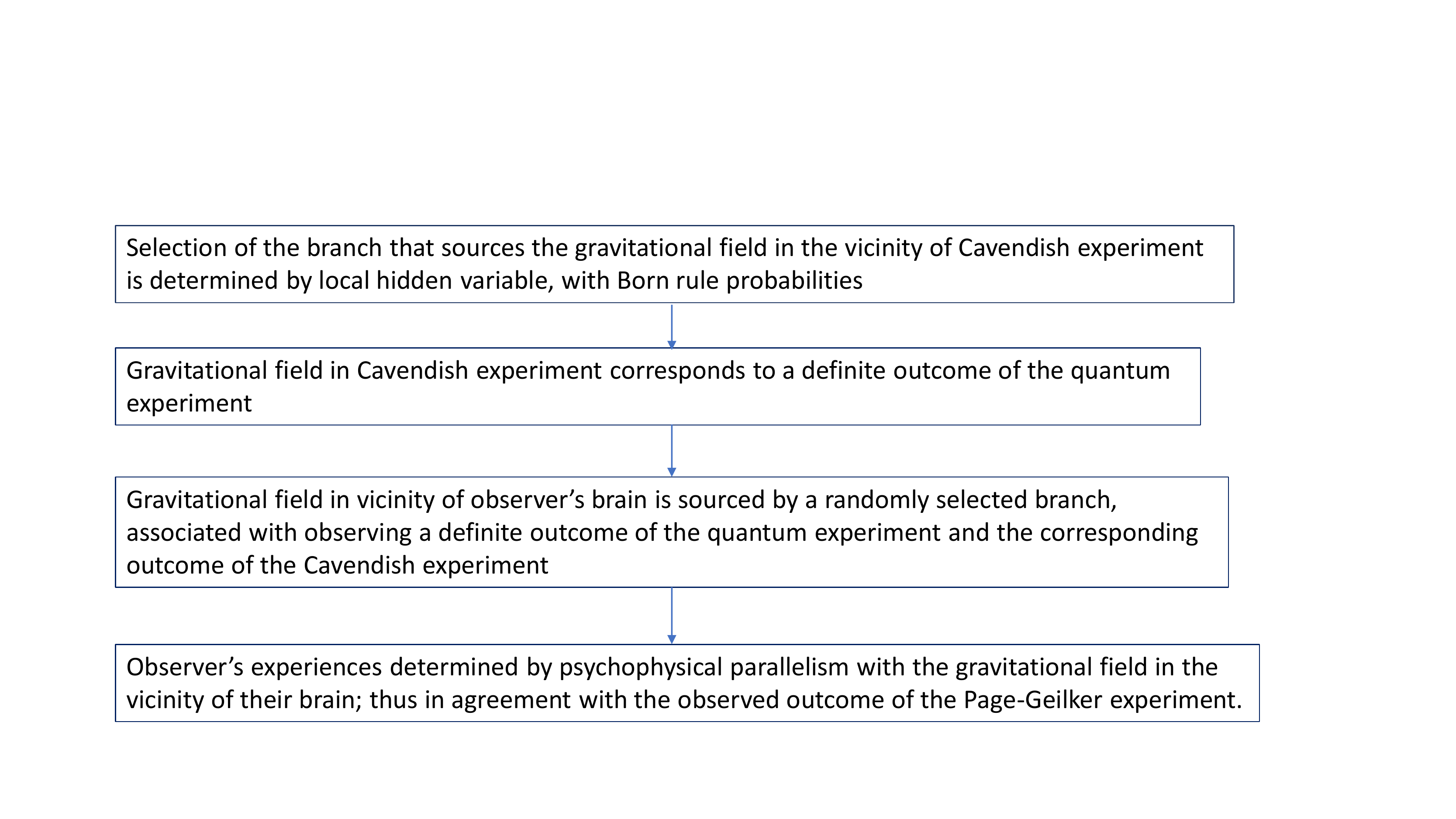} 
\caption{Flow chart summarising the chain of predictions from a 
local hybrid model for a Page-Geilker experiment.}
\label{hybridchart1}
\end{figure}

We can treat this as a large number
of parallel experiments, each observing the decay of an
unstable nucleus.   
The quantum predictions can easily be reproduced by a local hidden variable theory,
by assigning a variable corresponding to a gamma ray emission
time.   This gives us a local hidden variable model that
correctly predicts the probabilities of mass configurations. 
In a local hybrid model, the gravitational fields in the vicinity of 
Page and Geilker's Cavendish experiments may be 
determined by these local hidden variables.  
Assuming, as we must in such models, 
that the experimenters' perceptions are determined by gravitational fields, 
we obtain predictions in line with the observed outcomes.  (See
Figs. \ref{decay2}
and \ref{hybridchart1}.)

\subsection{Local hybrid models and experiments on
  entangled systems}

Consider now an experiment in which a singlet is created at a source,
and the two spin $1/2$ particles are propagated to widely separated
detectors
$D_A$ and $D_B$,
which carry out measurements about axes $\ul{a}$ and $\ul{b}$
chosen from CHSH pairs on each wing, with the choices made 
in near real time by local quantum random number generators.
Suppose that these measurement choices and measurement outcomes are
both amplified and recorded by 
placing macroscopic masses in one of four macroscopically distinct configurations,
depending on the choice and outcome, on each wing. 
Suppose further that the corresponding gravitational fields are
directly measured, ideally (for reasons we will discuss in more 
detail later) 
in a way that allows the entire process on each wing, from
the particles entering the detector to the gravitational field
measurement being completed and recorded, to take
place in space-like separated regions.   

As Bell showed, the measurement outcomes
predicted by any local hidden variable model 
respect Bell inequalities, while quantum 
theory predicts violations. 
In our hybrid model, the matter follows unitary quantum evolution,
and so, in any given Everettian branch, we expect violations of Bell
inequalities. 
However, according to the model,
the gravitational fields in the vicinity of $D_A$ and $D_B$,
which also effectively define measurement outcomes,  are
determined by local hidden variables, and so they {\it respect}
Bell inequalities.     

For a loophole free test, it is crucial that these gravitational fields are directly
measured and the measurements recorded and cross-correlated.   
To see this, note first that, although we have constrained and
interpreted the local
hybrid models so that the gravitational and matter degrees of freedom normally
appear to be aligned, this is not guaranteed in all circumstances.
In particular, it cannot be guaranteed in this experiment, since
the matter follows standard quantum evolution and violates Bell
inequalities, while the gravitational degrees of freedom are
determined by locally causal hidden variables and respect
Bell inequalities.   Observing that the matter violates Bell
inequalities is thus not enough to infer that the gravitational
field does.

\begin{figure}[h]
\centering
\includegraphics[width=\linewidth, height=10cm]{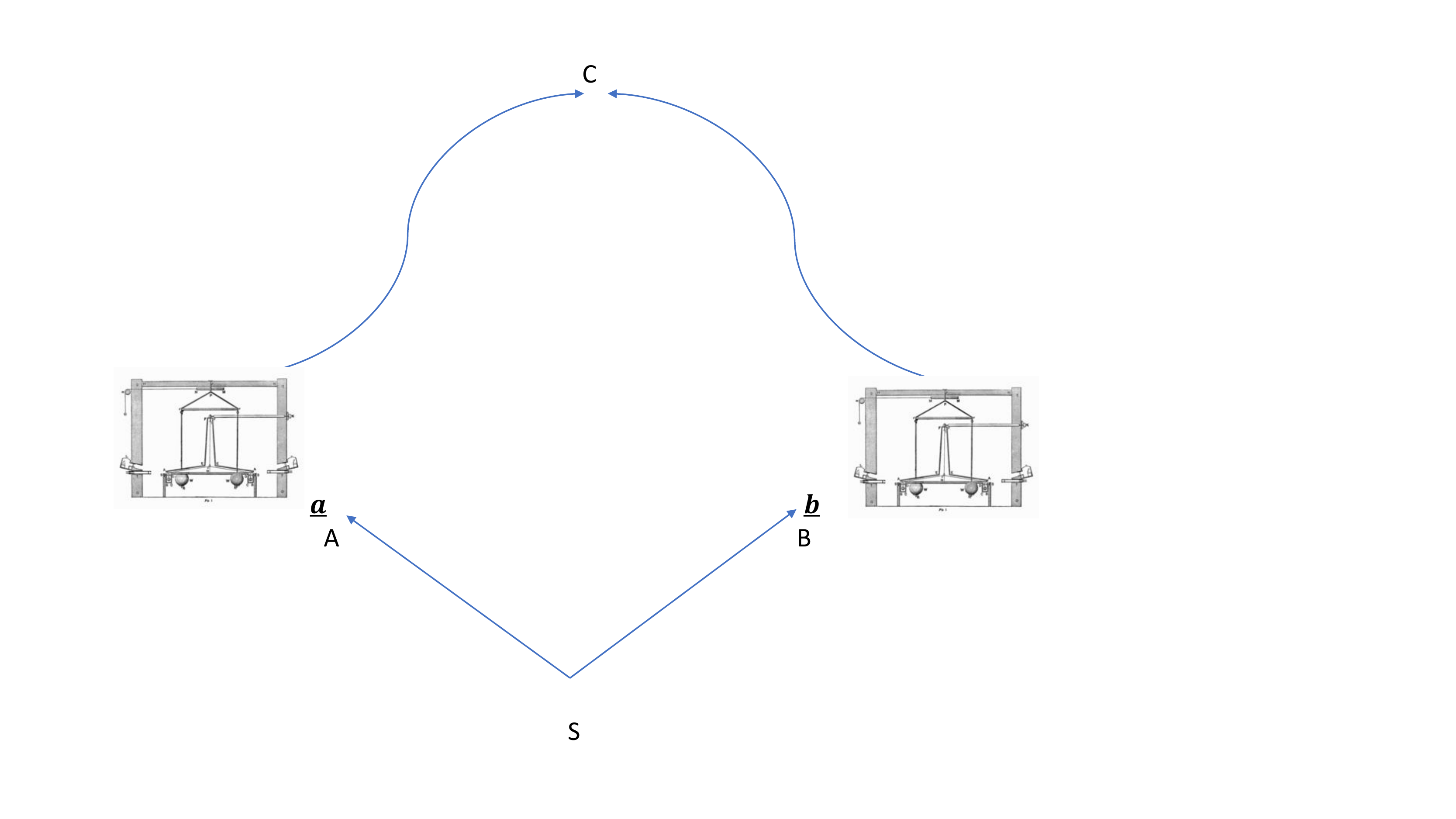} 
\caption{Schematic description of a direct test of the Bell
non-locality of the Newtonian gravitational field.   
A source S generates entangled particles, which are
sent to space-like separated regions.  
Local random choices of measurements A and B result
in outcomes a and b.   Both the choices and outcomes
are encoded in macroscopic mass configurations, whose
gravitational fields are directly measured by a 
Cavendish type experiment.   The outcomes of
these measurements are recorded, and the records
are sent and cross-correlated at some future point C.
The outcomes are recorded and cross-correlated in the matter
states, which are not directly correlated with perception
in this model.   However the gravitational field at C
is sourced by matter distribution arising from an Everett branch   
drawn randomly with Born weight, and so reflects the 
statistics of the gravitational field measurements at A and B.
If these respect Bell inequalities, so will the perceived
records at C.}
\label{gravbell}
\end{figure}

\begin{figure}[h]
\centering
\includegraphics[width=\linewidth, height=10cm]{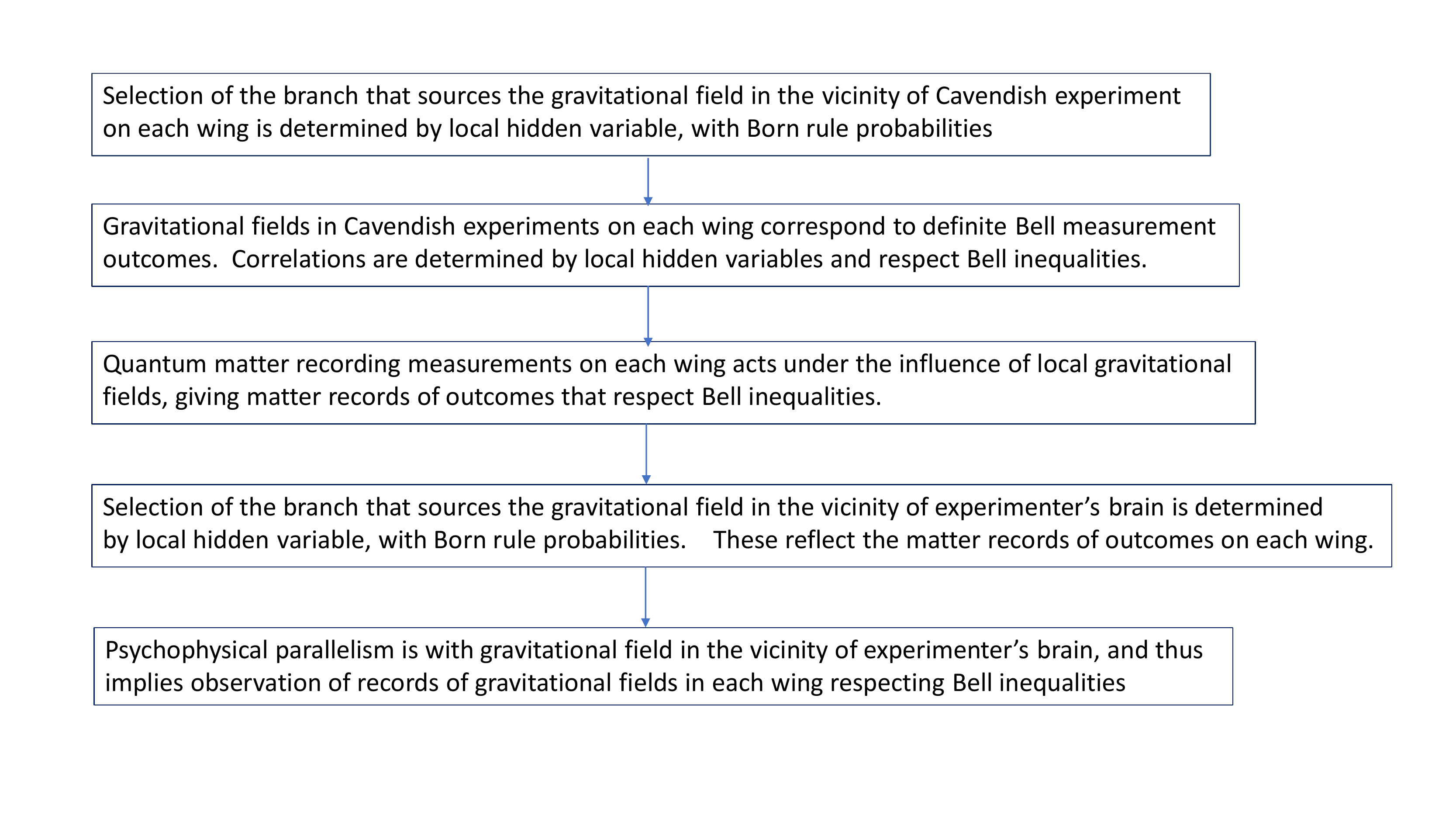} 
\caption{
Flow chart summarising the chain of predictions from a 
local hybrid model for an experiment directly testing
the Bell non-locality of the Newtonian gravitational field.
}
\label{hybridchart2}
\end{figure}

Now, by hypothesis, in these hybrid models, the perceptions and 
memories of observers at C are also correlated with
the local gravitational fields rather than with the Everettian matter state.
The local gravitational fields at C may in principle depend on physics
throughout the past light cone of C, and in particular on hidden
variables associated with both entangled particles at the source S.
However, we have the constraint that the probability distribution of
the gravitational fields at C reflects the Born weighted probability
distribution of mass configurations in the Everett branches there.
Measuring the gravitational fields at A and B and recording these
in the matter state thus ensures that the recorded and perceived
probability distribution at C respects Bell inequalities if 
the measurement outcomes at A and B did.  (See Figs. \ref{gravbell}
and \ref{hybridchart2}.)

This gives one specific motivation 
for a test of whether Bell nonlocality in
the gravitational field is directly observable.   
Such experiments were first proposed
in Refs. \cite{kent2005causal,kent2009proposed}, but without this specific added 
motivation.   These proposals were set in the
context of proposing a general intrinsic definition of 
Bell non-locality applicable to general space-times in
a large class of stochastic theories of gravity. 
While these (and potentially other) proposed definitions
remain interesting, they are not necessary in order to make the essential point:
we can test the Bell non-locality of the gravitational
field using measurements of Newtonian gravity in 
an approximately fixed and approximately flat space-time.  In
particular, we can carry out such tests on or around Earth.

\begin{figure}[h]
\centering
\includegraphics[width=\linewidth, height=10cm]{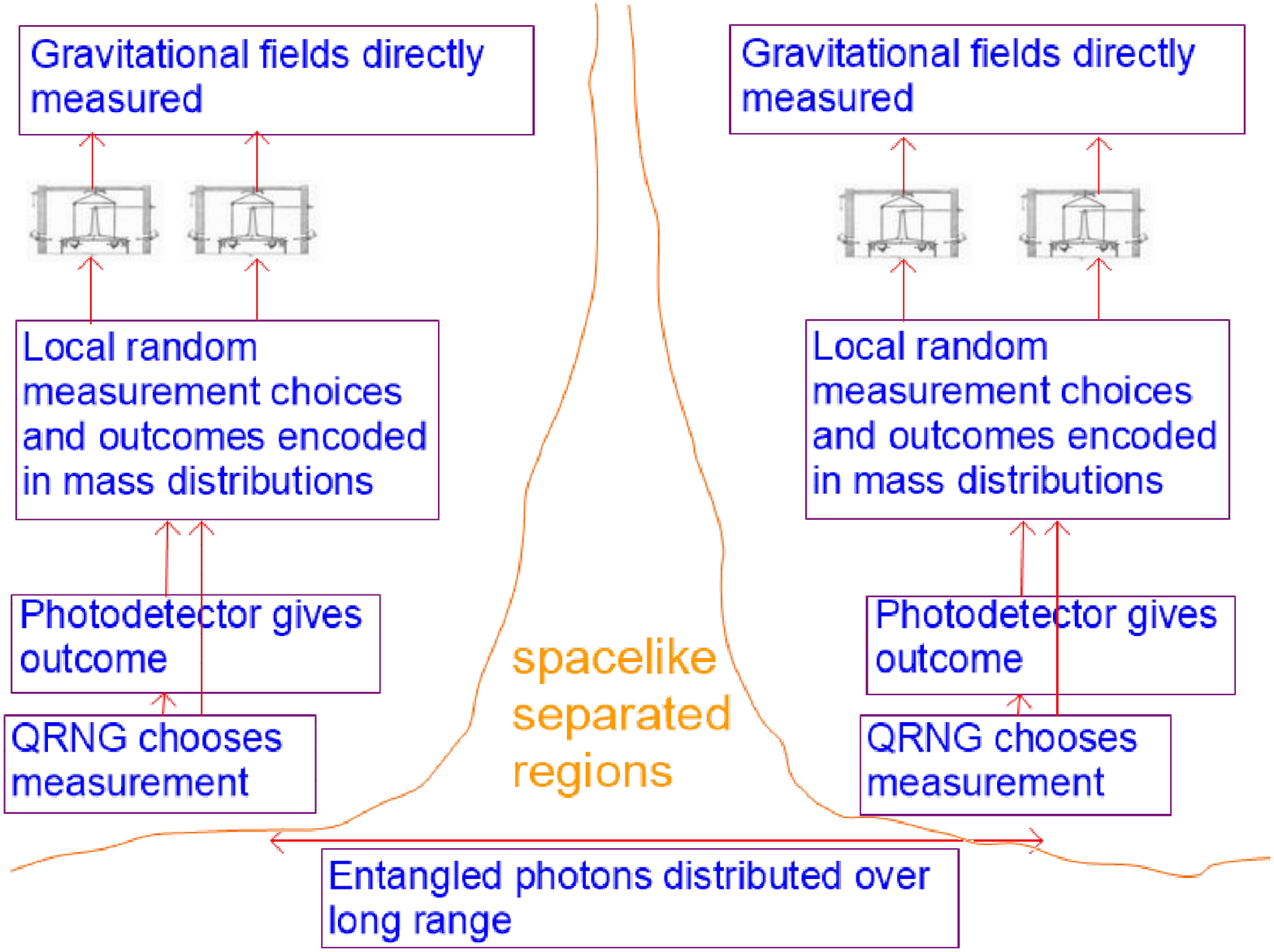} 
\caption{Schematic description of causality relations in direct test of the Bell
non-locality of the Newtonian gravitational field.   The entire
sequence of processes on the left and right wings must be carried
out in space-like separated regions.}
\label{gravbelltest}
\end{figure}

Like general relativity, hybrid models allow
information about the gravitational field in
a localized region to propagate everywhere within
its future light cone.   So it is crucial to ensure in such experiments
that the settings and outcomes on each wing all amplified into 
macroscopically distinct mass configurations, whose
gravitational fields are directly measured, with the
{\it entire processes} on each wing being space-like
separated.    These are very demanding requirements,
but the test is intriguing enough to justify 
exploring what is possible with technological ingenuity.
(See Fig. \ref{gravbelltest}.)

\section{Discussion}

Motivated by the example of semi-classical gravity, 
we have tried to explore the scope for other hybrid
models that combine unitary evolution of quantum
matter with a classical gravitational field.
We have found several problems.   
One possible response is that the idea may have been worth
looking into, but the problems turn out to be too daunting
and the possible solutions too unattractive
to pursue it further.   

Taking a more nuanced view, though, I think it worth reviewing
some possible objections case by case, since they
raise independently interesting questions. 

\begin{itemize}
\item   {\it To be testable in the forseeable future, hybrid models of
classical gravity and quantum matter need to predict that macroscopic mass distributions can
appear to be anomalously related to the Newtonian
gravitational field (or the metric), and this is not only contrary
to experience but also absurd.}

Semi-classical gravity illustrates that a hybrid theory, and one which
may yet turn out to be rigorously well defined, can predict
such anomalies.   Semi-classical gravity is indeed contradicted
by existing experiment, but a theory that predicts subtler anomalies
may not be.   One might reasonably assign Bayesian prior weights against
such models, because of a preference for quantum
gravity or on other theoretical grounds,
but I find it hard to see a case for making these weights
so close to dogmatic that experimental test is essentially
pointless.    Are we really so confident that unifying quantum
theory and gravity will produce nothing counter-intuitive?

\item  {\it Hybrid models with unitary quantum evolution can't
be right because Everettian quantum theory doesn't make sense.}

Hybrid models predict one classical space-time associated
with all the Everett branches.   This can only be consistent
with observation if psychophysical parallelism is with
the gravitational field rather than the quantum matter
state.   It is irrelevant here whether Everettian quantum
theory makes sense or not, or is attractive or not, in contexts
where the standard version of Everettian pyschophysical parallelism
(i.e. with the quantum matter) is used.    The relevant question is whether 
one is willing to consider the possibility that psychophysical
parallelism with the gravitational field could be a lawlike fact about the world.  

\item{\it Hybrid models with unitary quantum evolution
can't be right because Everettian quantum theory
makes perfect sense but makes completely different
predictions from those that are supposed to come
from hybrid models.}

See above.   Whatever predictions Everettian quantum theory
is supposed to make are contingent on psychophysical
parallelism being with the quantum matter state.
The first relevant question here is whether it is conceivable
that the quantum matter state could contain branches
that have the functional behaviour of observers 
interacting with an environment, without those
observers being conscious.  
One's answer to that should depend on one's
stance on the problem of consciousness, not on
one's stance on many-worlds quantum theory {\it per se}.
To those who argue for the logical conceivability
of philosophical zombies \cite{sep-zombies}, it is. 
To dogmatic functionalists \cite{sep-functionalism}, of course
it is not -- but then in my view, the problems of consciousness are so deep, 
with so many thoughtful people in different camps, that 
Aumann's theorem \cite{aumann1976agreeing} ought to deter anyone from dogmatism.   

The second relevant question is whether psychological
parallelism with the gravitational field could plausibly
work.   This is an interesting question, relevant not
just to the type of hybrid models we're considering, but
in principle to any discussion of quantum theory and
gravity.    It is particularly relevant, though, in
models where the gravitational and matter
degrees of freedom need not always be macroscopically
aligned, in which case it may be empirically testable. 
Tentatively, the answer seems to be that, although the
differences in the gravitational fields associated with 
different brain states are tiny, there presently seems no compelling reason
to reject the possibility that they are capable of
encoding the associated mind states.   

\item{\it Hybrid models in which a single spacetime
is defined globally by a randomly chosen Everettian branch
from the unitarily evolving quantum state face 
a consistency problem, because the branch structure
depends on the space-time, and vice versa.   
This may be worse than the back-reaction consistency problem
of semi-classical gravity.}

This may be correct.   Certainly there is no literature
addressing this problem.    One alternative is to 
consider models in which branches are chosen locally
according to rules determined by local physics. 

\item{\it Hybrid models in which a single spacetime
is defined locally by (inter alia) local hidden
variables are inelegant.   It is not
clear how to motivate any 
specific local hidden variable model, and no
specific proposal has been identified.}

Certainly, adding local hidden variables detracts
from the formal elegance exemplified by semi-classical gravity,
which motivated this exploration. 
It is also true that no specific model has been
identified, and there may be no compelling model.    
The same is true of local hidden variable
theories underlying quantum theory: no one
has ever presented a particularly compelling
local hidden variable theory.    Nonetheless,
the local hidden variable hypothesis has motivated
Bell experiments to exclude
local hidden variable theories as a class. 

Similarly, a positive feature of the present proposal
is that it can be excluded by experiment.
The relevant experiments present an interesting
technological challenge.    

\item{\it Hybrid models need not have gravitational field at $P$ determined
by an Everettian branch $B(P)$ chosen with Born weight probability.
Other probability distributions are logically possible.}

This is true, though the postulate seems a reasonable ansatz.
Unless we replace it by another constraint that has clear
empirical consequences, hybrid theories become essentially untestable.  

\end{itemize} 

\vskip10pt
\begin{acknowledgments}
This work was partially 
supported by
Perimeter Institute for Theoretical Physics. Research at Perimeter
Institute is supported by the Government of Canada through Industry
Canada and by the Province of Ontario through the Ministry of
Research and Innovation.   I thank Fay Dowker, Lucien Hardy
and Don Page for very helpful discussions.    
\end{acknowledgments}

\bibliographystyle{unsrtnat}
\bibliography{collapselocexpt}{}

\begin{thebibliography}{28}
\providecommand{\natexlab}[1]{#1}
\providecommand{\url}[1]{\texttt{#1}}
\expandafter\ifx\csname urlstyle\endcsname\relax
  \providecommand{\doi}[1]{doi: #1}\else
  \providecommand{\doi}{doi: \begingroup \urlstyle{rm}\Url}\fi

\bibitem[Eppley and Hannah(1977)]{eppley1977necessity}
Kenneth Eppley and Eric Hannah.
\newblock The necessity of quantizing the gravitational field.
\newblock \emph{Foundations of Physics}, 7\penalty0 (1-2):\penalty0 51--68,
  1977.

\bibitem[Mattingly(2006)]{mattingly2006eppley}
James Mattingly.
\newblock Why {E}ppley and {H}annah's thought experiment fails.
\newblock \emph{Physical Review D}, 73\penalty0 (6):\penalty0 064025, 2006.

\bibitem[Huggett and Callender(2001)]{huggett2001quantize}
Nick Huggett and Craig Callender.
\newblock Why quantize gravity (or any other field for that matter)?
\newblock \emph{Philosophy of Science}, 68\penalty0 (S3):\penalty0 S382--S394,
  2001.

\bibitem[Albers et~al.(2008)Albers, Kiefer, and
  Reginatto]{albers2008measurement}
Mark Albers, Claus Kiefer, and Marcel Reginatto.
\newblock Measurement analysis and quantum gravity.
\newblock \emph{Physical Review D}, 78\penalty0 (6):\penalty0 064051, 2008.

\bibitem[Kent(2018{\natexlab{a}})]{kent2018simple}
Adrian Kent.
\newblock Simple refutation of the {E}ppley-{H}annah argument.
\newblock \emph{arXiv preprint arXiv:1807.08708}, 2018{\natexlab{a}}.

\bibitem[Page and Geilker(1981)]{page1981indirect}
Don~N Page and CD~Geilker.
\newblock Indirect evidence for quantum gravity.
\newblock \emph{Physical Review Letters}, 47\penalty0 (14):\penalty0 979, 1981.

\bibitem[Kent(2005{\natexlab{a}})]{kent2005nonlinearity}
Adrian Kent.
\newblock Nonlinearity without superluminality.
\newblock \emph{Physical Review A}, 72\penalty0 (1):\penalty0 012108,
  2005{\natexlab{a}}.

\bibitem[Saunders et~al.(2010)Saunders, Barrett, Kent, and
  Wallace]{saunders2010many}
Simon Saunders, Jonathan Barrett, Adrian Kent, and David Wallace.
\newblock \emph{Many {W}orlds?: {E}verett, Quantum Theory, \& Reality}.
\newblock Oxford University Press, 2010.

\bibitem[Hu and Matacz(1995)]{hu1995back}
Bei-Lok Hu and Andrew Matacz.
\newblock Back reaction in semiclassical gravity: The {E}instein-{L}angevin
  equation.
\newblock \emph{Physical Review D}, 51\penalty0 (4):\penalty0 1577, 1995.

\bibitem[Flanagan and Wald(1996)]{flanagan1996does}
Eanna~E Flanagan and Robert~M Wald.
\newblock Does back reaction enforce the averaged null energy condition in
  semiclassical gravity?
\newblock \emph{Physical Review D}, 54\penalty0 (10):\penalty0 6233, 1996.

\bibitem[Martin and Verdaguer(1999)]{martin1999stochastic}
Rosario Martin and Enric Verdaguer.
\newblock Stochastic semiclassical gravity.
\newblock \emph{Physical Review D}, 60\penalty0 (8):\penalty0 084008, 1999.

\bibitem[Giulini and Kiefer(1995)]{giulini1995consistency}
Domenico Giulini and Claus Kiefer.
\newblock Consistency of semiclassical gravity.
\newblock \emph{Classical and Quantum Gravity}, 12\penalty0 (2):\penalty0 403,
  1995.

\bibitem[Hu and Phillip(2000)]{hu2000fluctuations}
BL~Hu and Nicholas~G Phillip.
\newblock Fluctuations of energy density and validity of semiclassical gravity.
\newblock \emph{International Journal of Theoretical Physics}, 39\penalty0
  (7):\penalty0 1817--1830, 2000.

\bibitem[Horowitz(1980)]{horowitz1980semiclassical}
Gary~T Horowitz.
\newblock Semiclassical relativity: The weak-field limit.
\newblock \emph{Physical Review D}, 21\penalty0 (6):\penalty0 1445, 1980.

\bibitem[Diosi(1987)]{diosi1987universal}
Lajos Diosi.
\newblock A universal master equation for the gravitational violation of
  quantum mechanics.
\newblock \emph{Physics Letters A}, 120\penalty0 (8):\penalty0 377--381, 1987.

\bibitem[Penrose(1996)]{penrose1996gravity}
Roger Penrose.
\newblock On gravity's role in quantum state reduction.
\newblock \emph{General Relativity and Gravitation}, 28\penalty0 (5):\penalty0
  581--600, 1996.

\bibitem[Tilloy and Di{\'o}si(2016)]{tilloy2016sourcing}
Antoine Tilloy and Lajos Di{\'o}si.
\newblock Sourcing semiclassical gravity from spontaneously localized quantum
  matter.
\newblock \emph{Physical Review D}, 93\penalty0 (2):\penalty0 024026, 2016.

\bibitem[Kirk(2015)]{sep-zombies}
Robert Kirk.
\newblock Zombies.
\newblock In Edward~N. Zalta, editor, \emph{The Stanford Encyclopedia of
  Philosophy}. Metaphysics Research Lab, Stanford University, summer 2015
  edition, 2015.

\bibitem[Levin(2018)]{sep-functionalism}
Janet Levin.
\newblock Functionalism.
\newblock In Edward~N. Zalta, editor, \emph{The Stanford Encyclopedia of
  Philosophy}. Metaphysics Research Lab, Stanford University, fall 2018
  edition, 2018.

\bibitem[Bell(1995)]{bell1995quantum}
John~S Bell.
\newblock Quantum mechanics for cosmologists.
\newblock In \emph{Quantum Mechanics, High Energy Physics And Accelerators:
  Selected Papers Of John S Bell (With Commentary)}, pages 793--819. World
  Scientific, 1995.

\bibitem[Kent(2005{\natexlab{b}})]{kent2005causal}
Adrian Kent.
\newblock Causal quantum theory and the collapse locality loophole.
\newblock \emph{Physical Review A}, 72\penalty0 (1):\penalty0 012107,
  2005{\natexlab{b}}.

\bibitem[Kent(2009)]{kent2009proposed}
Adrian Kent.
\newblock A proposed test of the local causality of spacetime.
\newblock In \emph{Quantum Reality, Relativistic Causality, and Closing the
  Epistemic Circle}, pages 369--378. Springer, 2009.

\bibitem[Aumann(1976)]{aumann1976agreeing}
Robert~J Aumann.
\newblock Agreeing to disagree.
\newblock \emph{The annals of statistics}, pages 1236--1239, 1976.

\bibitem[Page()]{pagepriv}
Don Page.
\newblock private communication.

\bibitem[Bassi et~al.(2010)Bassi, Deckert, and Ferialdi]{bassi2010breaking}
Angelo Bassi, D-A Deckert, and Luca Ferialdi.
\newblock Breaking quantum linearity: Constraints from human perception and
  cosmological implications.
\newblock \emph{EPL (Europhysics Letters)}, 92\penalty0 (5):\penalty0 50006,
  2010.

\bibitem[Kent(2018{\natexlab{b}})]{kent2018perception}
Adrian Kent.
\newblock Perception constraints on mass-dependent spontaneous localization.
\newblock \emph{arXiv preprint arXiv:1806.10396}, 2018{\natexlab{b}}.

\bibitem[Kent(2010)]{kent2010one}
Adrian Kent.
\newblock One world versus many: the inadequacy of {E}verettian accounts of
  evolution, probability, and scientific confirmation.
\newblock In Simon Saunders, Jonathan Barrett, Adrian Kent, and David Wallace,
  editors, \emph{Many {W}orlds?: {E}verett, Quantum Theory, \& Reality}, pages
  307--354. Oxford University Press, 2010.

\bibitem[Kent(2016)]{kent2016quanta}
Adrian Kent.
\newblock Quanta and qualia.
\newblock \emph{arXiv preprint arXiv:1608.04804}, 2016.

\end{thebibliography}
\end{document}